\documentclass[aps,prb,twocolumn,superscriptaddress,showkeys]{revtex4-1}

\usepackage[draft]{fixme}
\fxuselayouts{footnote}
\usepackage{amsmath}
\usepackage{amssymb}
\usepackage{graphicx}
\usepackage{color}
\usepackage{bm}
\usepackage{tikz}

\usepackage[normalem]{ulem} 

\newcommand{\diff}[2]{\frac{\partial #1}{\partial #2}}

\begin{document}

\title{Nonlocal response in thin-film waveguides: loss versus nonlocality and breaking of complementarity}
\date{\today}

\author{S\o ren Raza}
\affiliation{Department of Photonics Engineering, Technical University of Denmark, DK-2800 Kgs. Lyngby, Denmark}
\affiliation{Center for Electron Nanoscopy, Technical University of Denmark, DK-2800 Kgs. Lyngby, Denmark}
\author{Thomas Christensen}
\affiliation{Department of Photonics Engineering, Technical University of Denmark, DK-2800 Kgs. Lyngby, Denmark}
\affiliation{Center for Nanostructured Graphene, Technical University of Denmark, DK-2800 Kgs. Lyngby, Denmark}
\author{Martijn Wubs}
\affiliation{Department of Photonics Engineering, Technical University of Denmark, DK-2800 Kgs. Lyngby, Denmark}
\affiliation{Center for Nanostructured Graphene, Technical University of Denmark, DK-2800 Kgs. Lyngby, Denmark}
\author{Sergey I. Bozhevolnyi}
\affiliation{Institute of Sensors, Signal and Electrotechnics, University of Southern Denmark, DK-5230 Odense, Denmark}
\author{N. Asger Mortensen}
\email[]{asger@mailaps.org}
\affiliation{Department of Photonics Engineering, Technical University of Denmark, DK-2800 Kgs. Lyngby, Denmark}
\affiliation{Center for Nanostructured Graphene, Technical University of Denmark, DK-2800 Kgs. Lyngby, Denmark}

\begin{abstract}
We investigate the effects of nonlocal response on the surface-plasmon polariton guiding properties of the metal-insulator (MI), metal-insulator-metal (MIM), and insulator-metal-insulator (IMI) waveguides. The nonlocal effects are described by a linearized hydrodynamic model, which includes the Thomas--Fermi internal kinetic energy of the free electrons in the metal. We derive the nonlocal dispersion relations of the three waveguide structures taking into account also retardation and interband effects, and examine the delicate interplay between nonlocal response and absorption losses in the metal. We also show that nonlocality breaks the complementarity of the MIM and IMI waveguides found in the non-retarded limit.
\end{abstract}

\maketitle

\section{Introduction} \label{sec:intro}
\begin{figure}[b]
  \includegraphics[width=1\columnwidth]{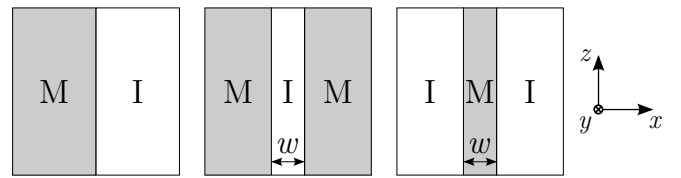}\\
  \caption{The three waveguide systems: metal-insulator (MI), metal-insulator-metal (MIM) and insulator-metal-insulator (IMI) along with the chosen coordinate system.}\label{fig:fig1}
\end{figure}

Guiding of light at metal-dielectric interfaces has attracted a lot of attention in recent years due to the subwavelength light confinement achievable by excitation of propagating surface-plasmon polariton (SPP) modes.~\cite{Zia:2006a, Berini:2009a} SPP guiding in a number of configurations is not limited by the diffraction limit, allowing for the manipulation and concentration of light on the nanoscale.~\cite{Gramotnev:2010} At the same time, stronger SPP mode confinement is typically associated with stronger mode absorption in metal, resulting in a trade-off between light confinement and propagation distances.~\cite{Maier:2007} This trade-off can be tailored by considering various waveguide structures, where especially waveguides based on thin metal films or narrow dielectric gaps between two metal surfaces have shown to provide a considerably better trade-off.~\cite{Bozhevolnyi:2006a, Sarid:1981} Symmetric metal-insulator-metal (MIM) and insulator-metal-insulator (IMI) waveguides, see Fig.~\ref{fig:fig1}, are the most fundamental of this class of waveguide structures, and provide a solid foundation for the understanding of more complex plasmonic waveguides. The key property of the IMI waveguide is its ability to support the so-called long-range SPP mode, which exhibits considerably low propagation loss. Furthermore, the MIM configuration forms the basis for the effective-index modeling (EIM) technique of more complex waveguiding structures, such as V-groove, slot and trench waveguides.~\cite{Bozhevolnyi:2006a, Hocker:1977}

The MIM and IMI waveguides have been extensively studied experimentally~\cite{Welford:1988,Bozhevolnyi:2005a,Dionne:2006a,Miyazaki:2006,Bozhevolnyi:2006b} and theoretically.~\cite{Sarid:1981, Burke:1986, Zia:2004a, Ginzburg:2006, Dionne:2006b} A key feature of any theoretical description of SPPs involves a suitable choice for the modeling of the response of free electrons of the metal. By far, the most common approach in the literature has been to apply the local-response approximation (LRA). The LRA solutions for the MIM and IMI structures were determined very early by Economou~\cite{Economou:1969a} and comprise two SPP modes being of even and odd symmetry with respect to the electric and magnetic fields, respectively. The properties of these modes are determined by their respective dispersion relations, i.e. by the relations between the frequency $\omega$ and the SPP propagation constant $k$, which are given by transcendental equations. In the nonretarded limit of the LRA, the surface modes of the MIM and IMI structures become identical,~\cite{Sernelius:2001} which is an interesting property that stems from Babinet's principle of complementary structures.~\cite{Rossouw:2012}

However, issues with the approach of the LRA arise when either considering large values of $k$, where an unphysical limit is found for the frequency, or considering narrow insulator or metal widths ($w<10$~nm), where singularities occur.~\cite{Maier:2007,Bozhevolnyi:2008a} In particular when investigating extremely narrow V-grooves, as recently realized experimentally,~\cite{Sondergaard:2012} with techniques such as EIM, the inadequacy of the LRA manifests itself. Nonlocal response (or spatial dispersion) has been shown to remove this flawed behavior of the SPP modes in waveguiding structures such as single metal-dielectric interfaces,~\cite{Boardman:1976a} infinite cylinders,~\cite{Aers:1980} and more recently, hyperbolic metamaterials,~\cite{Yan:2012a} conical tips,~\cite{Ruppin:2005,Wiener:2012} hybrid plasmonic waveguides,~\cite{Huang:2013a} wedges and V-grooves.~\cite{Toscano:2012c}

Extensive theoretical work has been done on the fundamental MIM and IMI waveguides, yet only a few studies~\cite{Boardman:1976a, Ruppin:2005b, Andersen:2012, Moreau:2013} have focused on nonlocal effects in these structures. In this paper, we fill this gap by determining the dispersion relations of SPP modes of the IMI and MIM waveguides taking into account nonlocal response, retardation effects as well as interband transitions in the metals. We also revisit the simple metal-insulator (MI) waveguide structure. The nonlocal response is described by a linearized semiclassical hydrodynamic model, \cite{Bloch:1933a, Boardman:1982a} which includes the quantum kinetics of the free-electron gas described by Thomas--Fermi theory.

The derivations and corresponding results for the nonlocal retarded dispersion relations of the IMI, MIM and MI waveguides are presented in Sec.~\ref{sec:theory}. With the dispersion relations for the three waveguides at hand, we examine in detail the interplay between losses and nonlocality in the metal by gradually increasing the absorption losses. The fundamental influence of losses versus nonlocality on the SPP dispersion has, to our knowledge, not yet been investigated. Furthermore, we compare modes of the IMI and MIM waveguides with and without retardation and nonlocality, and show that only in the non-retarded LRA do the modes of these two complementary waveguides become identical. Retardation and nonlocality are shown to break their complementarity. These topics are discussed in Sec.~\ref{sec:results}. Finally, Sec.~\ref{sec:conclusions} concludes the paper.

\section{Theory} \label{sec:theory}

\subsection{Nonlocal theory for thin-film systems} \label{sec:theoryA}
To determine the modes of thin-film systems, we first outline the main equations for the electric and magnetic fields that must be solved. We then consider the class of guided solutions with transverse magnetic (TM) polarization. The boundary conditions for the metal-dielectric interfaces are also discussed.

The free-electron gas of the metals comprising the thin-film waveguides is described by a nonlocal hydrodynamic equation of motion.~\cite{Boardman:1982a,Raza:2011} An intuitive way of describing the effect of nonlocal response is that it serves to smear out the charges at the surface of the metal on the scale of the Thomas--Fermi screening length.~\cite{David:2012,Pendry:2012} One of the key impacts of this charge smearing is the removal of field divergences that are known to occur in the LRA.~\cite{Toscano:2012,Fernandez-Dominguez:2012b,Ciraci:2012,Toscano:2012b,Fernandez-Dominguez:2012}
The hydrodynamic equation relating the current density $\mathbf{J}(\mathbf{r} ,\omega)$ to the electric field $\mathbf{E}(\mathbf{r},\omega)$ is given by~\cite{Raza:2011,Toscano:2012}
\begin{align}
    \frac{\beta_\textsc{f}^2}{\omega(\omega+i\gamma)} \bm \nabla \left[ \bm \nabla \cdot \mathbf{J}(\mathbf{r} ,\omega)\right] + \mathbf{J}(\mathbf{r} ,\omega) = \sigma(\omega) \mathbf{E}(\mathbf{r}, \omega), \label{eq:mainhydro}
\end{align}
where $\sigma(\omega) = i\varepsilon_0\omega_\text{p}^2/(\omega+i\gamma)$ is the Drude conductivity, and $\beta_\textsc{f}^2=(3/5) v_\textsc{f}^2$ is the nonlocal parameter obtained from Thomas--Fermi theory, where $v_\textsc{f}$ is the Fermi velocity of the metal. By combining Eq.~(\ref{eq:mainhydro}) with Maxwell's equations, the general equations describing the electric field $\mathbf{E}(\mathbf{r},\omega)$ in a metal with hydrodynamic nonlocal response can be compactly written as~\cite{Boardman:1976a,Raza:2011}
\begin{subequations}\label{eq:mainhydrogrp}
\begin{align}
    \left(\nabla^2 + k_\text{m}^2 \right) \bm \nabla \times \mathbf{E}(\mathbf{r},\omega) &= 0, \label{eq:mainET} \\
    \left(\nabla^2 + k_\text{nl}^2 \right) \bm \nabla \cdot \mathbf{E}(\mathbf{r},\omega) &= 0, \label{eq:mainEL}
\end{align}
\end{subequations}
where $k_\text{m} \equiv k_0 \sqrt{\varepsilon_\text{m}}$ is the usual wave vector in the metal while $k_\text{nl}\equiv \sqrt{\omega^2+i\gamma\omega-\omega_\text{p}^2/ \varepsilon_\infty} / \beta_\textsc{f}$ is the additional longitudinal wave vector present in a nonlocal description of the metal. Here, $k_0\equiv \omega/c$ is the vacuum wave vector, $\varepsilon_\text{m}\equiv \varepsilon_\infty(\omega) - \omega_\text{p}^2/(\omega^2+i\gamma\omega)$ is the local-response Drude permittivity including additional frequency-dependent polarization effects through $\varepsilon_\infty(\omega)$ not due to the free-electron plasma response.

The electric field in the insulator regions with permittivity $\varepsilon_\text{d}$ is described by the Helmholtz equation
\begin{align}
    \left(\nabla^2 + k_\text{d}^2 \right) \mathbf{E}(\mathbf{r},\omega) = 0, \label{eq:mainED}
\end{align}
where $k_\text{d}\equiv k_0 \sqrt{\varepsilon_\text{d}}$ is the wave vector in the insulator.

Once the electric field has been determined, the magnetic field $\mathbf{H}(\mathbf{r},\omega)$ can be found from Faraday's law
\begin{align}
    \mathbf{H}(\mathbf{r},\omega) = \frac{1}{i\omega\mu_0} \bm{\nabla} \times \mathbf{E}(\mathbf{r},\omega), \label{eq:mainH}
\end{align}
and then the free-electron current density $\mathbf{J}(\mathbf{r},\omega)$ in the metal can be found as
\begin{align}
    \mathbf{J}(\mathbf{r},\omega) = \bm{\nabla} \times \mathbf{H}(\mathbf{r},\omega) + i\omega\varepsilon_0 \varepsilon_\infty(\omega) \mathbf{E}(\mathbf{r},\omega). \label{eq:mainJ}
\end{align}

Without loss of generality we set the propagation direction along the $z$-axis and define the $x$-axis as perpendicular to the propagation plane, as in Fig.~\ref{fig:fig1}. Then the electric and magnetic fields for TM polarization can be simplified to
\begin{subequations}\label{eq:EHgrp}
\begin{align}
    \mathbf{E}(\mathbf{r},\omega) &= \left[E_x(x) \bm{\hat{e}_x} + E_z(x) \bm{\hat{e}_z} \right] e^{ikz}, \label{eq:Efield} \\
    \mathbf{H}(\mathbf{r},\omega) &= H_y(x) e^{ikz} \bm{\hat{e}_y}, \label{eq:Hfield}
\end{align}
\end{subequations}
where $k$ is the SPP propagation constant. With the definitions in Eq.~(\ref{eq:EHgrp}), we can simplify the general expressions of Eqs.~(\ref{eq:mainhydrogrp}-\ref{eq:mainJ}) to the following component form
\begin{subequations} \label{eq:nonlocalgrp}
\begin{align}
    &\left(\diff{^2}{x^2}-\kappa_\text{nl}^2 \right) \left[ k E_z(x) - i \diff{E_x(x)}{x} \right] = 0, \label{eq:mainEcomp1} \\
    &\left(\diff{^2}{x^2}-\kappa_\text{m}^2 \right) \left[ k E_x(x) + i \diff{E_z(x)}{x} \right] = 0, \label{eq:mainEcomp2} \\
    &H_y(x)=\frac{1}{\omega \mu_0} \left[ kE_x(x)+i\diff{E_z(x)}{x} \right], \label{eq:mainHcomp} \\
    &J_x(x) = -ikH_y(x) +i\omega\varepsilon_0 \varepsilon_\infty E_x(x), \label{eq:mainJcomp}
\end{align}
\end{subequations}
which are to be solved in the metal regions, while in the insulator regions the governing equations are
\begin{subequations} \label{eq:localgrp}
\begin{align}
    &\left(\diff{^2}{x^2}-\kappa_\text{d}^2 \right) E_x(x) = 0, \label{eq:mainEDcomp1} \\
    &E_z(x) = \frac{i}{k} \diff{E_x(x)}{x}. \label{eq:mainEDcomp2}
\end{align}
\end{subequations}
The magnetic fields in the insulator regions are also determined using Eq.~(\ref{eq:mainHcomp}). For conveninence, we have defined a propagation constant normal to the interfaces in the respective regions given as
\begin{align}
    \kappa_{j}^2 \equiv k^2-k_j^2 \qquad\text{for}\  j\in\{\mathrm{m},\mathrm{d},\mathrm{nl}\}. \label{eq:kappa}
\end{align}

With Eqs.~(\ref{eq:nonlocalgrp}) and (\ref{eq:localgrp}), solutions for the electric field, magnetic field and current density can be determined in the metal and insulator regions. At the metal-dielectric interfaces, we must connect the solutions using boundary conditions (BCs). Maxwell's BCs provide two of the three needed, namely the continuity of the tangential components of the electric and magnetic fields ($E_z$ and $H_y$, respectively). In our treatment, we neglect effects due to electron spill-out and quantum tunneling, which unambiguously determines the third and additional BC to be the vanishing of the normal component of the free-electron current density ($J_x$).~\cite{Jewsbury:1981,Boardman:1982a,Raza:2011,Yan:2012a} With this assumption we reduce our range of consideration to widths larger than $1\ \mathrm{nm}$ for the MIM waveguide.~\cite{Zuloaga:2009a,Esteban:2012}

\subsection{Single metal-insulator (MI) interface}
Before considering thin-film waveguides, it is instructive to revisit the fundamental problem of SPPs propagating at a single MI interface. The MI problem with hydrodynamic nonlocal response in the metal has been solved by Boardman \textit{et al.},~\cite{Boardman:1976a} in the simplest of cases where interband contributions and intraband damping were neglected. These results were recently generalized to include such contributions,~\cite{Moreau:2013} however, without considering the delicate interplay between the absorption losses and nonlocality in the metal, which we examine in Sec.~\ref{sec:results}.

The retarded nonlocal dispersion relation for a single MI interface is exactly given as
\begin{equation}
    1=-\frac{\varepsilon_\text{m} \kappa_\text{d}}{\varepsilon_\text{d} \kappa_\text{m}}-\delta_\text{nl}, \label{eq:MI}
\end{equation}
where $\delta_\text{nl}$ is an important nonlocal correction that will also appear below for the more complex thin-film waveguides, and is given as
\begin{equation}
    \delta_\text{nl}=\frac{k^2}{\kappa_\text{nl} \kappa_\text{m}}  \frac{\varepsilon_\text{m}-\varepsilon_\infty}{ \varepsilon_\infty}. \label{eq:deltaNL}
\end{equation}
We emphasize that when $\beta_\textsc{f} \rightarrow 0$, the local-response dispersion relation in Ref.~\onlinecite{Economou:1969a} is retrieved since $\delta_\text{nl} \rightarrow 0$. We also note that the $k$-solutions of the radical equation in Eq.~(\ref{eq:MI}) can be obtained analytically by the standard method for solving radical equations through squaring, and can be represented in terms of the solutions of a third-order polynomial.

For completeness, we also note that in the non-retarded limit $c \rightarrow \infty$, the nonlocal correction $\delta_\text{nl}$ simplifies to
\begin{equation}
    \delta_\text{nl}^\text{nr} \equiv \lim_{c\rightarrow\infty} \delta_{\text{nl}} =  \frac{k}{\kappa_\text{nl}} \frac{\varepsilon_\text{m}-\varepsilon_\infty}{ \varepsilon_\infty}, \label{eq:deltaNLnr}
\end{equation}
and the dispersion relation for a single MI interface Eq.~(\ref{eq:MI}) simplifies to
\begin{equation}
    1=-\frac{\varepsilon_\text{m}}{\varepsilon_\text{d} }-\delta_\text{nl}^\text{nr}. \label{eq:MInr}
\end{equation}
The non-retarded local-response dispersion relation is retrieved by letting $\delta_\text{nl}^\text{nr}\rightarrow 0$ in Eq.~(\ref{eq:MInr}).

\subsection{Metal-insulator-metal (MIM)}
The problem of determining the SPP modes of the MIM waveguide can, as in the LRA, be simplified by considering the even and odd modes separately. The symmetry considerations apply to the electric field. The exact retarded nonlocal dispersion relation for the fundamental, even mode is
\begin{subequations} \label{eq:MIMgrp}
\begin{equation}
    \tanh\left(\frac{\kappa_\text{d} w}{2} \right) = -\frac{\varepsilon_\text{d} \kappa_\text{m}}{\varepsilon_\text{m} \kappa_\text{d}} \left(1 + \delta_\text{nl}\right), \label{eq:MIMeven}
\end{equation}
while for the odd mode we find
\begin{equation}
    \coth\left(\frac{\kappa_\text{d} w}{2} \right) = -\frac{\varepsilon_\text{d} \kappa_\text{m}}{\varepsilon_\text{m} \kappa_\text{d}}\left(1 + \delta_\text{nl}\right), \label{eq:MIModd}
\end{equation}
\end{subequations}
where $w$ is the width of insulator slab. These equations are in agreement with the recent results.~\cite{Moreau:2013} In the nonretarded limit, Eq.~(\ref{eq:MIMgrp}) simplifies to
\begin{subequations} \label{eq:MIMnrgrp}
\begin{align}
    \tanh\left(\frac{k w}{2} \right) = -\frac{\varepsilon_\text{d}}{\varepsilon_\text{m}}\left(1 + \delta_\text{nl}^\text{nr}\right), \label{eq:MIMevennr} \\
    \coth\left(\frac{k w}{2} \right) = -\frac{\varepsilon_\text{d}}{\varepsilon_\text{m}}\left(1 + \delta_\text{nl}^\text{nr}\right). \label{eq:MIModdnr}
\end{align}
\end{subequations}

As previously mentioned, inclusion of nonlocal response regularizes the unphysical divergences encountered in the LRA. This property is also preserved for the MIM waveguide, and we may see how it comes about by examining the limit $w\rightarrow 0$ for the fundamental mode (which in the LRA produces a singularity). In this regard, we may neglect retardation effects and additionally simplify Eq.~(\ref{eq:MIMevennr}) by using the small-$x$ expansion $\tanh(x)\simeq x$. This yields
\begin{equation}
    k=-\frac{2\varepsilon_\text{d}}{\varepsilon_\text{m}} \left[w-\Delta_\textsc{mim} \right]^{-1}, \hspace{2mm} \Delta_\textsc{mim}=\frac{i\varepsilon_\text{d}(\varepsilon_\text{m}-\varepsilon_\infty)}{2k_\text{nl}\varepsilon_\text{m}\varepsilon_\infty} \label{eq:MIMklimit}
\end{equation}
where $\Delta_\textsc{mim}$ is the nonlocal correction, which vanishes in the local-response limit $\beta_\textsc{f} \rightarrow 0$. We emphasize that $k$ stays finite even in the case of $w=0$ in contrast to the diverging local-response relation~\cite{Bozhevolnyi:2008a} given by Eq.~(\ref{eq:MIMklimit}) with $\Delta_\textsc{mim}=0$.
\begin{figure*}[!tb]
  \includegraphics[width=1\textwidth]{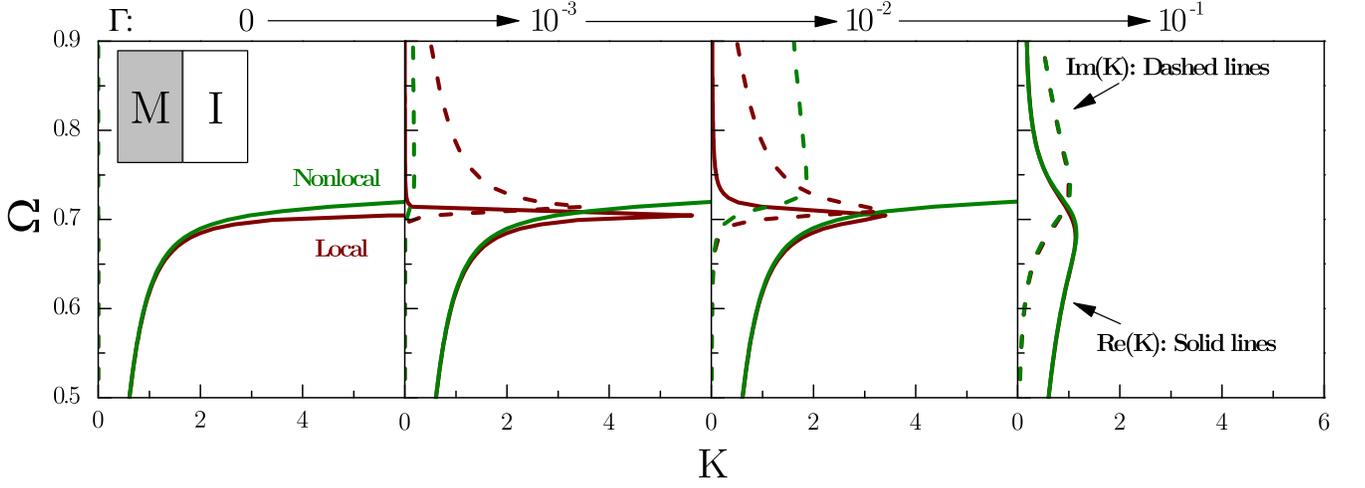}\\
  \caption{Local and nonlocal complex dispersion relations of the SPP mode of the MI waveguide with $\varepsilon_\text{d}=1$ and metal losses increasing from $\Gamma=0$ to $\Gamma=10^{-1}$. Local results are shown in red, while nonlocal are shown in green. Solid lines display the real part of the propagation constant, $\text{Re}(K)$, while the dashed lines display the imaginary part of the propagation constant, $\text{Im}(K)$. The value $\eta=5 \times 10^{-3}$ suitable for noble metals has been used.}
  \label{fig:fig2}
\end{figure*}

\subsection{Insulator-metal-insulator (IMI)}
The two SPP modes of the IMI waveguide can be classified into even and odd modes, as in the case of the MIM waveguide. However, while the symmetry characterization applied to the electric field for the MIM waveguide, here it is with respect to the magnetic field.~\cite{Yan:2012a} The nonlocal modes for the IMI waveguide have previously been studied in the case of a lossless metal without interband contributions.~\cite{Ruppin:2005b} Here, we generalize these results to include such contributions, which are important in realistic waveguides. The retarded nonlocal dispersion relation for the odd and even modes are
\begin{subequations} \label{eq:IMIgrp}
\begin{align}
    \coth \left(\frac{\kappa_\text{m} w}{2} \right) &= -\frac{ \varepsilon_\text{m} \kappa_\text{d} }{ \varepsilon_\text{d} \kappa_\text{m} } - \delta_\text{nl} \coth \left(\frac{\kappa_\text{nl} w}{2}\right), \label{eq:IMIodd} \\
    \tanh \left(\frac{\kappa_\text{m} w}{2} \right) &= -\frac{ \varepsilon_\text{m} \kappa_\text{d} }{ \varepsilon_\text{d} \kappa_\text{m} } - \delta_\text{nl} \tanh \left(\frac{\kappa_\text{nl} w}{2} \right). \label{eq:IMIeven}
\end{align}
\end{subequations}
We note that, in contrary to the MIM waveguide, the odd mode Eq.~(\ref{eq:IMIodd}) is in fact the fundamental mode.

In the nonretarded limit, Eqs.~(\ref{eq:IMIodd}) and (\ref{eq:IMIeven}) simplify to
\begin{subequations} \label{eq:IMInrgrp}
\begin{align}
    \coth \left(\frac{k w}{2} \right) &= -\frac{ \varepsilon_\text{m}}{ \varepsilon_\text{d}} - \delta_\text{nl}^\text{nr} \coth \left(\frac{\kappa_\text{nl} w}{2}\right), \label{eq:IMIoddnr} \\
    \tanh \left(\frac{k w}{2} \right) &= -\frac{ \varepsilon_\text{m} }{ \varepsilon_\text{d} } - \delta_\text{nl}^\text{nr} \tanh \left(\frac{\kappa_\text{nl} w}{2} \right). \label{eq:IMIevennr}
\end{align}
\end{subequations}

As in the case for the MIM waveguide, we can again examine the limit of $w\rightarrow 0$ for the fundamental mode. Neglecting retardation effects and using the small-$x$ expansion $\coth(x) \simeq 1/x$, we find
\begin{subequations} \label{eq:IMIklimit}
\begin{align}
    &k=-\frac{2\varepsilon_\text{d}}{w \varepsilon_\text{m}} \left[1-\left(\frac{\Delta_\textsc{imi}}{w}\right)^2 \right], \\
    &\Delta_\textsc{imi}= \frac{2\varepsilon_\text{d}}
    {k_\text{nl}\varepsilon_\text{m}} \sqrt{\frac{\varepsilon_\text{m}-\varepsilon_\infty}{\varepsilon_\text{m} \varepsilon_\infty}}.
\end{align}
\end{subequations}
Here, $\Delta_\textsc{imi}$ is the nonlocal correction for the IMI waveguide that vanishes for vanishing $\beta_\textsc{f}$. Unlike the MIM waveguide, the nonlocal correction does not regularize the diverging $k$ when $w=0$. Due to the confinement of the electron plasma in the IMI waveguide, as opposed to the MIM waveguide, the regularization of the dispersion likely requires inclusion of electron spill-out, which is not treated here, see Sec.~\ref{sec:theoryA}. Further elaboration on the comparison of Eq.~(\ref{eq:IMIklimit}) with Eq.~(\ref{eq:MIMklimit}), in the context of complementarity, is done in Sec.~\ref{sec:compl}.

\section{Results} \label{sec:results}
The dispersion relations introduced in Sec.~\ref{sec:theory} are complex-valued transcendental equations of the implicit form $F(\omega, k)=0$, with the propagation constant in general being a complex number $k=k'+ik''$. Thus to determine the waveguide modes, solutions to the dispersion relations must be found in the complex $k$-plane for each frequency, which in general is a non-trivial task. Fortunately, a robust and reliable numerical scheme suitable for determining the zeros in the complex plane, based on the Cauchy integral formula, has been previously developed~\cite{Delves:1967,Chen:2000} and is employed in this work.

In the following, we focus on the free-electron properties (i.e.\ $\varepsilon_\infty=1$) of the modes of the three different waveguides. This allows us to rescale the dispersion relations with normalized quantities, here introduced as $\Omega = \omega/\omega_\text{p}$, $K=kc/\omega_\text{p}$, $\Gamma = \gamma/\omega_\text{p}$, $\eta = \beta_\textsc{f}/c$, and for the IMI and MIM waveguides, $W=w\omega_\text{p}/c$. The normalized parameters $\Gamma$ and $\eta$ characterize the losses and the strength of nonlocality in the metals, respectively.
\begin{figure*}[!tb]
  \includegraphics[width=1\textwidth]{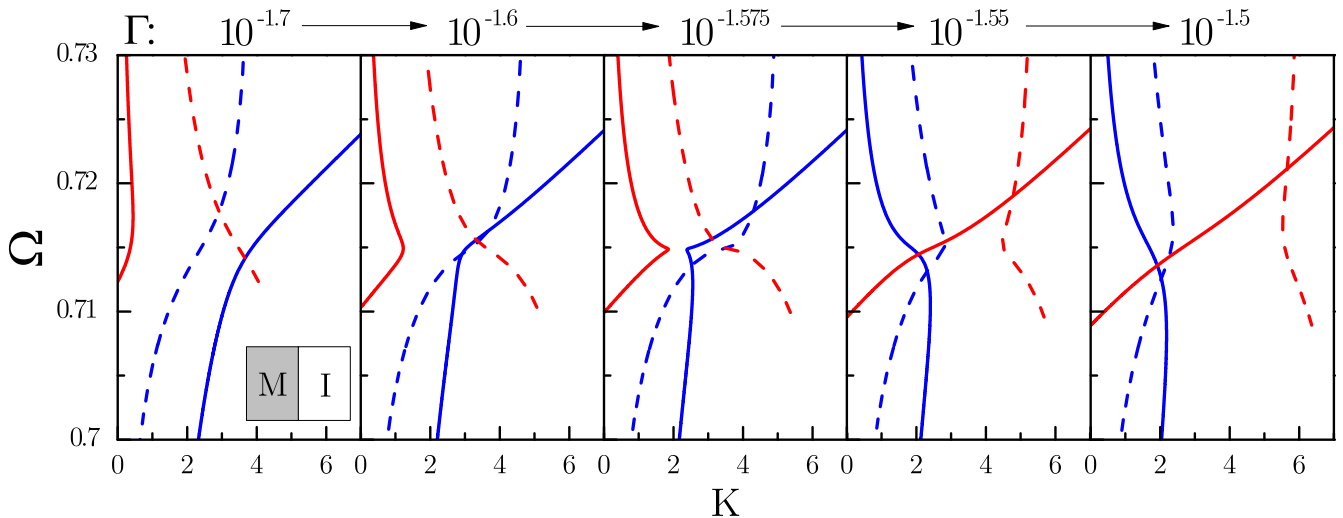}\\
  \caption{Nonlocal complex dispersion relations of the SPP mode of the MI waveguide with $\varepsilon_\text{d}=1$ and metal losses increasing from $\Gamma=10^{-1.8}$ to $\Gamma=10^{-1.6}$. Solid lines display the real part of the propagation constant, $\text{Re}(K)$, while the dashed lines display the imaginary part of the propagation constant, $\text{Im}(K)$. The transition and mode-evolution from nonlocality to loss-dominated behavior is explored. The value $\eta=5 \times 10^{-3}$ suitable for noble metals has been used.}
  \label{fig:fig2b}
\end{figure*}

This section is divided into two parts: Sec.~\ref{sec:lossvsnl} concerns the interplay between metal losses and nonlocality. Here, we first study this interplay in the simple MI waveguide that does not contain any geometric length scales, whereafter we examine how nonlocal effects are enhanced in confined waveguides such as the MIM and IMI waveguides. Sec.~\ref{sec:compl} deals with the breaking of complementarity in the MIM and IMI waveguides due to nonlocal response.

\subsection{Losses vs nonlocality} \label{sec:lossvsnl}

\paragraph*{MI waveguide:} Due to the absence of length scales associated with the geometry, the MI waveguide is an ideal system to study, when considering the interplay between losses and nonlocality of the SPP mode. An additional benefit of studying the MI structure is that it is not obscured by the effects of multiple interface reflections that is present in the MIM and IMI structures, such that only the intrinsic properties of free electrons affect the waveguiding properties.

The interplay between losses and nonlocality in the MI waveguide is seen in Fig.~\ref{fig:fig2}, where we display the effect of increasing the metal losses on the local and nonlocal dispersion relations of the SPP mode, given by Eq.~(\ref{eq:MI}) with $\eta=5 \times 10^{-3}$. In the lossless case ($\Gamma=0$), the local dispersion relation converges towards the well-known $\Omega_\textsc{sp}=1/\sqrt{2 \varepsilon_\text{d}}$ limit for large $K$ values, while the nonlocal dispersion relation increases in frequency without bound, in agreement with earlier results.~\cite{Boardman:1976a} However, in the presence of very weak losses ($\Gamma=10^{-3}$) the infinite $K$ values at the frequency $\Omega_\textsc{sp}$ in the LRA are removed and the SPP mode bends back. This back-bending effect is a well-known textbook result,~\cite{Maier:2007} which occurs for any positive value for $\Gamma$ in the LRA. The extreme sensitivity to even minute losses in the LRA is due to the vanishing group velocity $v_\text{g}=\partial\omega/\partial k$ at $\Omega_\textsc{sp}$.~\cite{Pedersen:2008} In striking contrast, the nonlocal SPP mode [i.e.\ $\text{Re}(K)$] is robust due to the finite group velocity $v_\text{g} \geq \beta_\textsc{f}$. Consequently, no pronounced slow-light enhancement of weak losses takes place and the nonlocal SPP mode does not bend back until the losses of the system start to dominate. Although non-zero $\text{Im}(K)$ is generated for the nonlocal SPP mode for $\Gamma \neq 0$, the real part of the propagation constant $\text{Re}(K)$ remains largely unaffected. It is also interesting to note that the behaviour of $\text{Im}(K)$, which is related to the SPP propagation length $l_\textsc{spp}$ through $l_\textsc{spp}=1/[2\text{Im}(K)]$, changes drastically from $\Gamma=10^{-3}$ to $\Gamma=10^{-2}$. For $\Gamma=10^{-3}$ the nonlocal SPP mode propagates longer than the local one in the frequency region $\Omega>\Omega_\textsc{sp}$, while the opposite result is seen for $\Gamma=10^{-2}$. At the same time $\text{Re}(K)$ for the nonlocal mode is unchanged and substantially larger than in the LRA, resulting in shorter wavelengths and thereby stronger confinement of the SPP mode at the MI surface. Not until $\Gamma=10^{-1}$, which is significantly larger than the nonlocal parameter $\eta$, do the losses in the metal dominate over nonlocality and force the nonlocal SPP mode to bend back. At such losses, the local and nonlocal models result in almost identical solutions.

The transition of the nonlocal mode from being dominated primarily by nonlocality to being dominated by losses (i.e. $\Gamma=10^{-2} \rightarrow 10^{-1}$ in Fig.~\ref{fig:fig2}) is investigated in more detail in Fig.~\ref{fig:fig2b}. To explain the transition we must also consider the presence of the Brewster mode (for clarity not shown in Fig.~\ref{fig:fig2}) and not only the SPP mode. In Fig.~\ref{fig:fig2b} we see the merging of two separated modes, plotted as red and blue lines. For the lowest loss of $\log(\Gamma)=-1.7$, the red line corresponds to the continuation of the Brewster mode to frequencies lower than $\Omega = 1$, which in the lossless case would be a forbidden region (i.e. only purely lossy solutions exist).~\cite{Novotny:2012} The blue line represents the standard, low-loss, nonlocal SPP mode. As the losses increase [$\log(\Gamma) = -1.6 \rightarrow -1.575$], the real parts of the dispersion of the Brewster mode and SPP mode begin to merge. At approximately $\log(\Gamma) = -1.55$ the mode-appearance has qualitatively changed, with the appearance of the usual well-known loss-dominated SPP mode (in blue), which is also present in LRA, as well as the emergence of a relatively flat band, nonlocal surface plasmon mode (in red) near the surface plasmon resonance $\Omega_\textsc{sp}$. We notice that the nonlocal flat band mode is significantly damped in comparison with the usual SPP mode, and that the damping increases drastically with increased material loss. In contrast, the usual SPP mode is not nearly so sensitive to the small change in material loss from $\log(\Gamma) = -1.55$ to $\log(\Gamma) = -1.5$.

We now present a simple analysis to understand when the metal losses dominate nonlocal effects in the MI waveguide. The back-bending occurs at the frequency $\Omega_\textsc{sp}$, where the propagation constant is significantly larger than the free-space propagation constant. We can therefore justify to examine the simpler non-retarded dispersion relation given by Eq.~(\ref{eq:MInr}) instead of the retarded dispersion relation [Eq.~(\ref{eq:MI})]. From Eq.~(\ref{eq:MInr}), we see that nonlocality becomes negligible when $|\delta_\text{nl}^\text{nr}| \ll |1+\varepsilon_\text{m}/\varepsilon_\text{d}|$. Evaluating this condition at the SPP frequency with $\varepsilon_\text{d}=1$ (as in Fig.~\ref{fig:fig2}) for small $\Gamma$ leads to the simple condition for loss-dominated behavior
\begin{equation}
    \Gamma \gg \eta, \label{eq:MIloss}
\end{equation}
which is consistent with our numerical analysis. We point out that the loss parameter $\Gamma$ is just one of several options for introducing an imaginary part to the metal permittivity. An alternative approach to introducing losses is by simply adding a constant imaginary part to the lossless free-electron Drude model. In either case, the metal permittivity becomes complex-valued. To bridge these different approaches, we can relate the condition in Eq.~(\ref{eq:MIloss}) to the imaginary part of the metal permittivity by noting that $\text{Im}\left[\varepsilon_\text{m}\left(\Omega=\Omega_\textsc{sp}\right)\right] = 2\sqrt{2}\Gamma/\left(1+2\Gamma^2\right)$ in which case Eq.~(\ref{eq:MIloss}) can be rewritten as
\begin{equation}
    \text{Im}(\varepsilon_\text{m}) \gg \frac{v_\textsc{f}}{c}. \label{eq:MIloss2}
\end{equation}
In noble metals the nonlocal parameter is of the order $v_\textsc{f}/c \approx 10^{-3}$ and the losses are of the order $\text{Im}(\varepsilon_\text{m}) \approx 10^{0}$, which in general means that metal losses largely dominate nonlocal effects in the SPP mode of the MI waveguide.~\cite{Rakic:1998a}

\begin{figure}[!bt]
  \includegraphics[width=1\columnwidth]{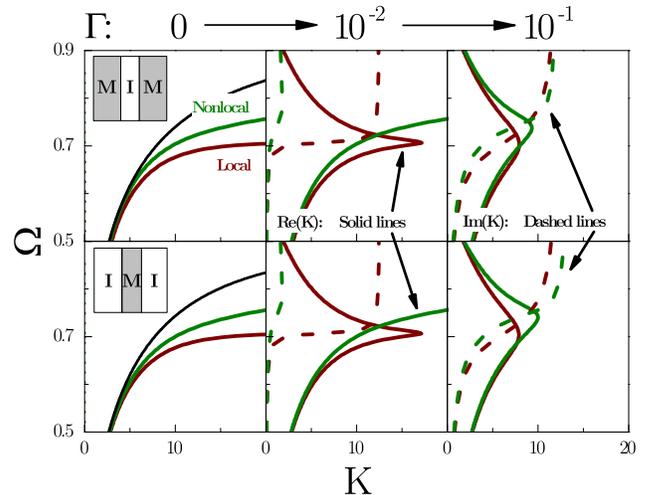}\\
  \caption{Local and nonlocal complex dispersion relations of the fundamental mode of the MIM (first row) and IMI (second row) waveguides with $W=0.25$, $\varepsilon_\text{d}=1$, $\eta=5 \times 10^{-3}$ and metal losses increasing from $\Gamma=0$ to $\Gamma=10^{-1}$. Local results are shown in red, while nonlocal are shown in green. Solid lines display the real part of the propagation constant $\text{Re}(K)$, while the dashed lines display the imaginary part of the propagation constant $\text{Im}(K)$. The black lines represent the approximate nonlocal dispersion relations given by Eqs.~(\ref{eq:MIMklimit}) and (\ref{eq:IMIklimit}) for the MIM and IMI waveguides, respectively.}
  \label{fig:fig3}
\end{figure}

\paragraph*{MIM and IMI waveguides:} Figure~\ref{fig:fig3} displays the effect of increasing losses on nonlocality for the fundamental modes of the MIM (first row) and IMI waveguides (second row) given by Eqs.~(\ref{eq:MIMeven}) and (\ref{eq:IMIodd}), respectively. The normalized width of the waveguides is set to $W=0.25$ which corresponds to a width of $w \approx 5~\text{nm}$ for Ag and Au ($\omega_\text{p} \approx 9~\text{eV}$). Considering the MIM waveguide first, we see that in the lossless case nonlocal response within the hydrodynamic model predicts a blueshift compared to the LRA (for a fixed $K$). As the losses in the metal increase ($\Gamma=10^{-2}$), the local dispersion relation [$\text{Re}(K)$] immediately bends back and the propagation length is significantly shorter than for the nonlocal case. Both of these effects are similar to those observed for the MI waveguide. When $\Gamma=10^{-1}$ the nonlocal dispersion relation also bends back and the nonlocal propagation length becomes comparable to LRA, albeit for $\Omega < \Omega_\textsc{sp}$ nonlocal response gives rise to longer propagation lengths than in the LRA. Although the nonlocal dispersion relation bends back at these large losses, nonlocal response still reveals a blueshift and larger values of $\text{Re}(K)$ than in the LRA.

The trend is very similar for the IMI waveguide (see second row of Fig.~\ref{fig:fig3}). In fact, in the LRA the difference between the fundamental modes of the IMI and MIM waveguides is practically negligible. As for the nonlocal case, the biggest difference between the IMI and MIM waveguides is seen for $\Gamma=10^{-1}$, where nonlocal response shows a slight increase in the maximum values of both the $\text{Re}(K)$ and $\text{Im}(K)$ for the IMI waveguide.

In Fig.~\ref{fig:fig3} we have also examined the validity of the approximate relations for the nonlocal fundamental modes of the MIM and IMI waveguides given by Eqs.~(\ref{eq:MIMklimit}) and (\ref{eq:IMIklimit}), respectively. They are plotted as black lines for the lossless case. We see that the approximate relations are in excellent agreement with the exact calculations when $KW \lesssim 1$.

The important feature for both waveguides is that even for large losses (of order $\Gamma=10^{-1}$) the nonlocal and local dispersion relations are different, in contrast to the MI waveguide. The nonlocal dispersion relations show larger values of $\text{Re}(K)$ than in the LRA for both waveguides. Thus, the limitations and undesired properties of metal losses are counteracted by nonlocality, which gives rise to a shorter wavelength of the SPP mode and thereby an increase of the mode confinement. These interesting features arise due to the multiple reflections present in the IMI/MIM waveguides, introducing a new length scale given by the scaled width of the slab $W$. The importance of nonlocal effects increases with decreasing width (or in general, size),~\cite{Raza:2011} and it is clear from the nonlocal dispersion relations for the IMI and MIM waveguides that the strength of nonlocality is different in these two waveguide structures, as also observed in Fig.~\ref{fig:fig3}. This difference arises due to the presence of confined nonlocal pressure waves, which are naturally only present in the IMI waveguide.

\subsection{Breaking of complementarity due to nonlocal response} \label{sec:compl}
\begin{figure}[!tb]
  \includegraphics[width=1\columnwidth]{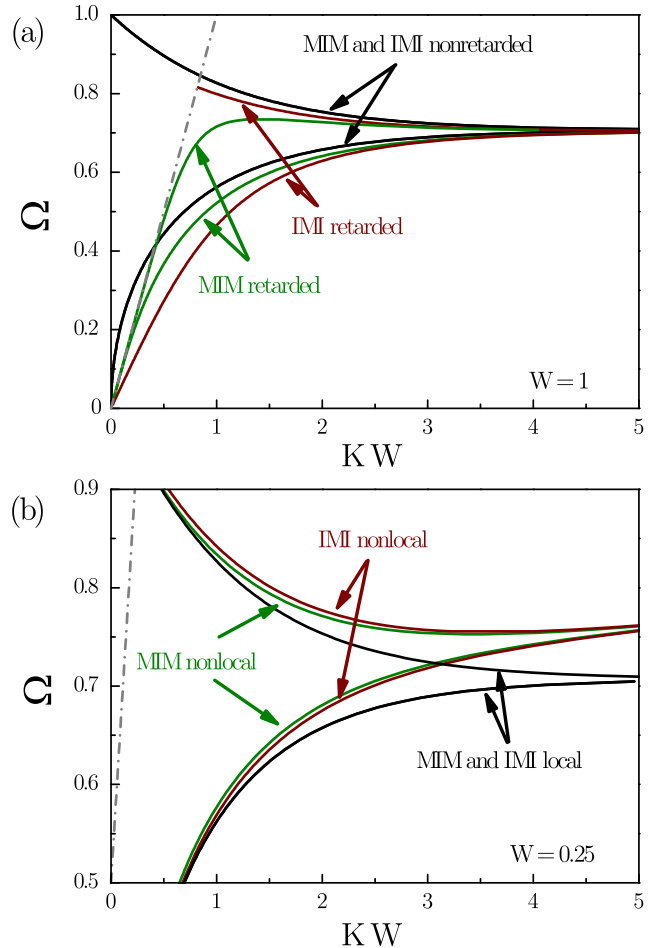}\\
  \caption{Plots of both surface modes of the lossless IMI and MIM waveguides (a) with and without retardation in the LRA and (b) with and without nonlocal response ($\eta=5\times10^{-3}$) in the non-retarded limit. The light line is shown in grey. The widths of the waveguides are (a) $W=1$ and (b) $W=0.25$.}
  \label{fig:fig4}
\end{figure}
It is well known that the LRA dispersion relations for the SPP modes of the MIM and IMI waveguides are identical in the non-retarded limit,~\cite{Maier:2007,Sernelius:2001} which is also clear from comparing Eq.~(\ref{eq:MIMnrgrp}) to Eq.~(\ref{eq:IMInrgrp}) with $\delta^\text{nr}_\text{nl}=0$. This property of identical surface modes in complementary waveguide structures, such as the MIM and IMI waveguides, is broken when retardation effects are included in the LRA, which become important for SPP propagation values $K$ close to the light line $K_0=\Omega$.~\cite{Sernelius:2001} Here, we show explicitly that nonlocal response also breaks the symmetry by considering the SPP modes of the MIM and IMI waveguides in the non-retarded limit, i.e.\ in the limit where $K\gg K_0$. In the following, we divide the discussion of breaking of complementarity into two parts: one due to retardation effects alone in the LRA, and one solely due to nonlocal response in the non-retarded limit. For the latter, we consider the non-retarded limit to ensure that the breaking of complementarity is due to nonlocal response rather than being attributed to retardation.

Breaking of complementarity is illustrated in Fig.~\ref{fig:fig4}. In Fig.~\ref{fig:fig4}(a), we plot the SPP modes of the MIM and IMI waveguides only in the LRA, displaying the effect of retardation. First, we note, as already mentioned, that the SPP modes of the MIM and IMI modes in the non-retarded limit are completely identical and overlap in Fig.~\ref{fig:fig4}(a) (black lines). When retardation effects are included the MIM (green lines) and IMI (red lines) surface modes are no longer identical for $K$-values close to the light line. The main consequence of properly taking retardation into account is that no guided modes exist above the light line (grey line). In Fig.~\ref{fig:fig4}(a), we clearly see that the retarded modes terminate at the light line, unlike the non-retarded modes.

Figure~\ref{fig:fig4}(b) shows the nonlocal and local SPP modes of the MIM and IMI waveguides calculated in the non-retarded limit. We see clearly that nonlocal response distinguishes between the MIM and IMI waveguide modes, for both of the two surface modes. This effect was observed earlier,~\cite{Abajo:2008a} but not elaborated on. As the propagation constant increases both of the nonlocal modes of both waveguides convergence towards the hydrodynamic nonlocal large-$K$ limit $K=\Omega/\eta$, as for the MI waveguide,~\cite{Boardman:1976a,Yan:2012a} and become indistinguishable. Finally, we also note the characteristic shift to higher frequencies of both nonlocal SPP modes compared to the LRA.

The breaking of the complementarity property of the MIM and IMI waveguides due to nonlocal response can of course be understood from the fact that the dispersion relations for the two waveguides are different even in the non-retarded limit, as seen by comparison of Eq.~(\ref{eq:MIMnrgrp}) with Eq.~(\ref{eq:IMInrgrp}). We can quantify this difference for the fundamental mode of the two structures by considering the difference $\Delta k = k_\textsc{mim} - k_\textsc{imi}$, where $k_\textsc{mim}$ and $k_\textsc{imi}$ are given by the approximate relations in Eqs.~(\ref{eq:MIMklimit}) and (\ref{eq:IMIklimit}), respectively. Using a Pad\'{e} approximation, we find to the lowest order in $w$ that
\begin{align}
    \Delta k \simeq -\frac{2\varepsilon_\text{d}}{\varepsilon_\text{m}} \frac{\Delta_\textsc{imi}^2}{w^3}. \label{eq:deltak}
\end{align}
From Eq.~(\ref{eq:deltak}) we clearly see that in the absence of nonlocal response ($\Delta_\textsc{imi}=0$), the difference between the fundamental modes of the MIM and IMI waveguides vanishes. Additionally, we observe that the dispersion-difference depends strongly on the width. At very narrow widths, we therefore expect a strong breaking of complementarity. Thus, in the presence of nonlocality, a thin film of electron gas embedded in an insulator behaves qualitatively different from a thin insulator gap embedded in an electron gas.

In an intuitive, but simplified picture of nonlocality, one could be inclined to attribute the complementarity breaking to the nonlocal smearing of the induced surface charge. In the nonlocal hydrodynamic model the induced surface charge is smeared over a length scale comparable to the Thomas--Fermi screening length, leading to an effective width increase (decrease) for the MIM (IMI) waveguide. In this picture the dispersion relations of the nonlocal IMI and MIM SPP modes should then be below and above the local dispersion relations, respectively, which is not the case, see Fig.~\ref{fig:fig4}(b). In fact the nonlocal IMI and MIM SPP modes are always above the local results, discrediting the simple interpretation of nonlocal response as local response with effective size parameters.

The complementarity breaking originates from the inclusion of pressure waves in the description of a metal with nonlocal response. More precisely, the breaking is due to the confinement of these pressure waves in the IMI waveguide, which becomes more important for narrower widths. This confinement results in a significantly different description of the IMI waveguide compared to the MIM waveguide, where the pressure waves are not confined. For this reason, nonlocal effects are also stronger in the IMI waveguide, as can be seen from the presence of only the nonlocal IMI correction in Eq.~(\ref{eq:deltak}). In contrast in the non-retarded LRA, the absence of both retardation and the pressure waves leads to a faulty identical treatment of the MIM and IMI waveguides.

\section{Conclusions} \label{sec:conclusions}
The effects of nonlocal response, described by a linearized hydrodynamic model, on the waveguiding properties of the MI, MIM, and IMI waveguides have been investigated. The corresponding dispersion relations for the three waveguides have been derived, taking into account nonlocality, interband transitions, and retardation. The intriguing transition from nonlocal- to loss-dominated waveguiding behavior, which has not previously been studied extensively, was examined for the MI system, demonstrating that nonlocal response can counteract the effects of low metal losses. In the LRA the presence of even minute losses drastically alters the dispersion relation of the SPP mode due the slow-light regime at the surface plasmon frequency $\Omega_\textsc{sp}$. For larger losses, the effects of nonlocality in the MI structure is less important, and the difference between local and nonlocal response becomes negligible. In general, for the MI structure, the impact of metal losses is much more pronounced than that of nonlocal effects, partially due to the high losses in metals and partially due to the absence of any geometric length scale in the MI structure.

Conversely, for the MIM and IMI structures, the presence of an additional length scale, given by the geometric width of the waveguide, yields a comparative boost to the effect of nonlocality vis-\`{a}-vis the effect of metal losses. In turn, the increased strength of nonlocality gives rise to larger propagation constants and thereby an increased plasmonic confinement of the SPP modes. Nonlocal effects are shown to be slightly stronger in the IMI waveguide due to the presence of confined longitudinal pressure waves, which are absent in the MIM structure.

Lastly, we also examined the complementarity property of the MIM and IMI waveguides in the context of Babinet's principle. In the non-retarded limit of the LRA, the waveguide modes of the MIM and IMI modes are known to be identical. When retardation is taken into account, this symmetry is broken. In addition, we have shown that in the nonretarded limit the symmetry is also broken by the inclusion of nonlocal effects due to the presence of nonlocal pressure waves.

\emph{Acknowledgments.}
The Center for Nanostructured Graphene is sponsored by the Danish National Research Foundation, Project DNRF58. The A. P. M{\o}ller and Chastine Mc-Kinney M{\o}ller Foundation is gratefully acknowledged for the contribution toward the establishment of the Center for Electron Nanoscopy.


\begin{thebibliography}{10}

\bibitem{Zia:2006a}
R. Zia, J.~A. Schuller, A. Chandran, and M.~L. Brongersma, {\em Plasmonics: the
  next chip-scale technology}, Mater. Today {\bf 9},  20  (2006).

\bibitem{Berini:2009a}
P. Berini, {\em Long-range surface plasmon polaritons}, Adv. Opt. Phot. {\bf
  1},  484  (2009).

\bibitem{Gramotnev:2010}
D.~K. Gramotnev and S.~I. Bozhevolnyi, {\em Plasmonics beyond the diffraction
  limit}, Nat. Photonics {\bf 4},  83  (2010).

\bibitem{Maier:2007}
S.~A. Maier, {\em Plasmonics: Fundamentals and Applications} (Springer, New
  York, 2007).

\bibitem{Bozhevolnyi:2006a}
S. Bozhevolnyi, {\em Effective-index modeling of channel plasmon polaritons},
  Opt. Express {\bf 14},  9467  (2006).

\bibitem{Sarid:1981}
D. Sarid, {\em Long-range surface-plasma waves on very thin metal films}, Phys.
  Rev. Lett. {\bf 47},  1927  (1981).

\bibitem{Hocker:1977}
G.~B. Hocker and W.~K. Burns, {\em Mode dispersion in diffused channel
  waveguides by the effective index method}, Appl. Optics {\bf 16},  113
  (1977).

\bibitem{Welford:1988}
K.~R. Welford and J.~R. Sambles, {\em Coupled surface plasmons in a symmetric
  system}, J. Mod. Opt. {\bf 35},  1467  (1988).

\bibitem{Bozhevolnyi:2005a}
S.~I. Bozhevolnyi, V.~S. Volkov, E. Devaux, and T.~W. Ebbesen, {\em Channel
  plasmon-polariton guiding by subwavelength metal grooves}, Phys. Rev. Lett.
  {\bf 95},  046802  (2005).

\bibitem{Dionne:2006a}
J.~A. Dionne, H.~J. Lezec, and H.~A. Atwater, {\em Highly confined photon
  transport in subwavelength metallic slot waveguides}, Nano Lett. {\bf 6},
  1928  (2006).

\bibitem{Miyazaki:2006}
H.~T. Miyazaki and Y. Kurokawa, {\em Squeezing visible light waves into a
  3-nm-thick and 55-nm-long plasmon cavity}, Phys. Rev. Lett. {\bf 96},  097401
   (2006).

\bibitem{Bozhevolnyi:2006b}
S.~I. Bozhevolnyi, V.~S. Volkov, E. Devaux, J.-Y. Laluet, and T.~W. Ebbesen,
  {\em Channel plasmon subwavelength waveguide components including
  interferometers and ring resonators}, Nature {\bf 440},  508  (2006).

\bibitem{Burke:1986}
J.~J. Burke, G.~I. Stegeman, and T. Tamir, {\em Surface-polariton-like waves
  guided by thin, lossy metal films}, Phys. Rev. B {\bf 33},  5186  (1986).

\bibitem{Zia:2004a}
R. Zia, M.~D. Selker, P.~B. Catrysse, and M.~L. Brongersma, {\em Geometries and
  materials for subwavelength surface plasmon modes}, J. Opt. Soc. Am. A {\bf
  21},  2442  (2004).

\bibitem{Ginzburg:2006}
P. Ginzburg, D. Arbel, and M. Orenstein, {\em Gap plasmon polariton structure
  for very efficient microscale-to-nanoscale interfacing}, Opt. Lett. {\bf 31},
   3288  (2006).

\bibitem{Dionne:2006b}
J.~A. Dionne, L.~A. Sweatlock, H.~A. Atwater, and A. Polman, {\em Plasmon slot
  waveguides: Towards chip-scale propagation with subwavelength-scale
  localization}, Phys. Rev. B {\bf 73},  035407  (2006).

\bibitem{Economou:1969a}
E.~N. Economou, {\em Surface plasmons in thin films}, Phys. Rev. {\bf 182},
  539  (1969).

\bibitem{Sernelius:2001}
B.~E. Sernelius, {\em Surface Modes in Physics} (Wiley-VCH, Berlin, 2001).

\bibitem{Rossouw:2012}
D. Rossouw and G.~A. Botton, {\em Resonant optical excitations in complementary
  plasmonic structures}, Opt. Express {\bf 20},  6968  (2012).

\bibitem{Bozhevolnyi:2008a}
S.~I. Bozhevolnyi and J. Jung, {\em Scaling for gap plasmon based waveguides},
  Opt. Express {\bf 16},  2676  (2008).

\bibitem{Sondergaard:2012}
T. S{\o}ndergaard, S.~M. Novikov, T. Holmgaard, R.~L. Eriksen, J. Beermann, Z.
  Han, K. Pedersen, and S.~I. Bozhevolnyi, {\em {Plasmonic black gold by
  adiabatic nanofocusing and absorption of light in ultra-sharp convex
  grooves}}, Nat. Commun. {\bf 3},  1  (2012).

\bibitem{Boardman:1976a}
A.~D. Boardman, B.~V. Paranjape, and Y.~O. Nakamura, {\em Surface
  plasmon-polaritons in a spatially dispersive inhomogeneous media}, Phys.
  Stat. Sol. (b) {\bf 75},  347  (1976).

\bibitem{Aers:1980}
G.~C. Aers, A.~D. Boardman, and B.~V. Paranjape, {\em Non-radiative surface
  plasmon-polariton modes of inhomogeneous metal circular-cylinders}, J. Phys.
  F: Met. Phys. {\bf 10},  53  (1980).

\bibitem{Yan:2012a}
W. Yan, M. Wubs, and N.~A. Mortensen, {\em Hyperbolic metamaterials: nonlocal
  response regularizes broadband super-singularity}, Phys. Rev. B {\bf 86},
  205429  (2012).

\bibitem{Ruppin:2005}
R. Ruppin, {\em Effect of non-locality on nanofocusing of surface plasmon field
  intensity in a conical tip}, Phys. Lett. A {\bf 340},  299   (2005).

\bibitem{Wiener:2012}
A. Wiener, A.~I. Fern{\'a}ndez-Dom{\'i}nguez, A.~P. Horsfield, J.~B. Pendry,
  and S.~A. Maier, {\em Nonlocal effects in the nanofocusing performance of
  plasmonic tips}, Nano Lett. {\bf 12},  3308  (2012).

\bibitem{Huang:2013a}
Q. Huang, F. Bao, and S. He, {\em Nonlocal effects in a hybrid plasmonic
  waveguide for nanoscale confinement}, Opt. Express {\bf 21},  1430  (2013).

\bibitem{Toscano:2012c}
G. Toscano, S. Raza, W. Yan, C. Jeppesen, S. Xiao, M. Wubs, A.-P. Jauho, S.~I.
  Bozhevolnyi, and N.~A. Mortensen, {\em Nonlocal response in plasmonic
  waveguiding with extreme light confinement}, arXiv:1212.4925  (2012).

\bibitem{Ruppin:2005b}
R. Ruppin, {\em Non-local optics of the near field lens}, J. Phys. Condens.
  Matter {\bf 17},  1803  (2005).

\bibitem{Andersen:2012}
K. Andersen, K.~W. Jacobsen, and K.~S. Thygesen, {\em Spatially resolved
  quantum plasmon modes in metallic nano-films from first-principles}, Phys.
  Rev. B {\bf 86},  245129  (2012).

\bibitem{Moreau:2013}
A. Moreau, C. Cirac\`{i}, and D.~R. Smith, {\em Impact of nonlocal response on
  metallodielectric multilayers and optical patch antennas}, Phys. Rev. B {\bf
  87},  045401  (2013).

\bibitem{Bloch:1933a}
F. Bloch, {\em Bremsverm\text{\"{o}}gen von atomen mit mehreren elektronen}, Z.
  Phys. A {\bf 81},  363  (1933).

\bibitem{Boardman:1982a}
A. Boardman, {\em Electromagnetic Surface Modes. Hydrodynamic theory of
  plasmon-polaritons on plane surfaces.} (John Wiley and Sons, Chichester,
  1982).

\bibitem{Raza:2011}
S. Raza, G. Toscano, A.-P. Jauho, M. Wubs, and N.~A. Mortensen, {\em Unusual
  resonances in nanoplasmonic structures due to nonlocal response}, Phys. Rev.
  B {\bf 84},  121412(R)  (2011).

\bibitem{David:2012}
C. David and F.~J. Garc\'{i}a~de Abajo, {\em Spatial nonlocality in the optical
  response of metal nanoparticles}, J. Phys. Chem. C {\bf 115},  19470  (2012).

\bibitem{Pendry:2012}
J.~B. Pendry, A. Aubry, D.~R. Smith, and S.~A. Maier, {\em Transformation
  optics and subwavelength control of light}, Science {\bf 337},  549  (2012).

\bibitem{Toscano:2012}
G. Toscano, S. Raza, A.-P. Jauho, N.~A. Mortensen, and M. Wubs, {\em Modified
  field enhancement and extinction in plasmonic nanowire dimers due to nonlocal
  response}, Opt. Express {\bf 20},  4176   (2012).

\bibitem{Fernandez-Dominguez:2012b}
A.~I. Fern\'{a}ndez-Dom\'{i}nguez, P. Zhang, Y. Luo, S.~A. Maier, F.~J.
  Garc\'{i}a-Vidal, and J.~B. Pendry, {\em Transformation-optics insight into
  nonlocal effects in separated nanowires}, Phys. Rev. B {\bf 86},  241110
  (2012).

\bibitem{Ciraci:2012}
C. Cirac\`{i}, R.~T. Hill, J.~J. Mock, Y. Urzhumov, A.~I.
  Fern\'{a}ndez-Dom\'{i}nguez, S.~A. Maier, J.~B. Pendry, A. Chilkoti, and
  D.~R. Smith, {\em Probing the ultimate limits of plasmonic enhancement},
  Science {\bf 337},  1072  (2012).

\bibitem{Toscano:2012b}
G. Toscano, S. Raza, S. Xiao, M. Wubs, A.-P. Jauho, S.~I. Bozhevolnyi, and
  N.~A. Mortensen, {\em Surface-enhanced Raman spectroscopy: nonlocal
  limitations}, Opt. Lett. {\bf 37},  2538  (2012).

\bibitem{Fernandez-Dominguez:2012}
A.~I. Fern\'{a}ndez-Dom\'{i}nguez, A. Wiener, F.~J. Garc\'{i}a-Vidal, S.~A.
  Maier, and J.~B. Pendry, {\em Transformation-optics description of nonlocal
  effects in plasmonic nanostructures}, Phys. Rev. Lett. {\bf 108},  106802
  (2012).

\bibitem{Jewsbury:1981}
P. Jewsbury, {\em Electrodynamic boundary conditions at metal interfaces}, J.
  Phys. F: Met. Phys. {\bf 11},  195  (1981).

\bibitem{Zuloaga:2009a}
J. Zuloaga, E. Prodan, and P. Nordlander, {\em Quantum description of the
  plasmon resonances of a nanoparticle dimer}, Nano Lett. {\bf 9},  887
  (2009).

\bibitem{Esteban:2012}
R. Esteban, A.~G. Borisov, P. Nordlander, and J. Aizpurua, {\em Bridging
  quantum and classical plasmonics with a quantum-corrected model}, Nat.
  Commun. {\bf 3},  825  (2012).

\bibitem{Delves:1967}
L.~M. Delves and J.~N. Lyness, {\em A numerical method for locating the zeros
  of an analytic function}, Math. Comput. {\bf 21},  543  (1967).

\bibitem{Chen:2000}
C. Chen, P. Berini, D. Feng, S. Tanev, and V.~P. Tzolov, {\em Efficient and
  accurate numerical analysis of multilayer planar optical waveguides in lossy
  anisotropic media}, Opt. Express {\bf 7},  260  (2000).

\bibitem{Pedersen:2008}
J.~G. Pedersen, S. Xiao, and N.~A. Mortensen, {\em Limits of slow light in
  photonic crystals}, Phys. Rev. B {\bf 78},  153101  (2008).

\bibitem{Novotny:2012}
L. Novotny and B. Hecht, {\em Principles of nano-optics}, 2 ed. (Cambridge
  University Press, Cambridge, 2012).

\bibitem{Rakic:1998a}
A.~D. Raki\'{c}, A.~B. Djuri\v{s}i\'{c}, J.~M. Elazar, and M.~L. Majewski, {\em
  {Optical properties of metallic films for vertical-cavity optoelectronic
  devices}}, Appl. Optics {\bf 37},  5271  (1998).

\bibitem{Abajo:2008a}
F.~J. Garci\'{a}~de Abajo, {\em Nonlocal effects in the plasmons of strongly
  interacting nanoparticles, dimers, and waveguides}, J. Phys. Chem. C {\bf
  112},  17983  (2008).

\end{thebibliography}
\end{document}